\documentclass[%
 aip,apl,
 amsmath,amssymb,
 reprint,%
]{revtex4-1}

\usepackage{graphicx}
\usepackage{dcolumn}
\usepackage{bm}

\usepackage[utf8]{inputenc}
\usepackage[T1]{fontenc}
\usepackage{mathptmx}

\usepackage[dvipsnames,svgnames,x11names]{xcolor} 
\definecolor{GreenMarker}{rgb}{0.0, 0.75, 0.0}

\definecolor{BlueMarker}{rgb}{0.0, 0.0, 255}

\begin{document}


\title[]{Enhancing the effective critical current density in a Nb superconducting thin film by cooling in an inhomogeneous magnetic field}

\author{D. A. D. Chaves}
\author{I. M. de Araújo}
\affiliation{ 
Departamento de Física, Universidade Federal de São Carlos, 13565-905 São Carlos, SP, Brazil
}%
\author{D. Carmo}
\affiliation{ 
Departamento de Física, Universidade Federal de São Carlos, 13565-905 São Carlos, SP, Brazil
}%
\affiliation{
Laboratório Nacional de Luz Síncrotron, Centro Nacional de Pesquisa em Energia e Materiais, 13083-100, Campinas, SP, Brazil
}%
\author{F. Colauto}
\affiliation{ 
Departamento de Física, Universidade Federal de São Carlos, 13565-905 São Carlos, SP, Brazil
}%
\author{A. A. M. de Oliveira}
\affiliation{%
Instituto Federal de Educação, Ciência e Tecnologia de São Paulo, Campus São Carlos, 13565-905, São Carlos, SP, Brazil
}%
\author{A. M. H. de Andrade}
\affiliation{%
Instituto de Física, Universidade Federal do Rio Grande do Sul, 91501-970 Porto Alegre, RS, Brazil
}%
\author{T. H. Johansen}
\affiliation{%
Department of Physics, University of Oslo, P.O. Box 1048 Blindern, 0316 Oslo, Norway
}%
\author{A. V. Silhanek}
\affiliation{Experimental Physics of Nanostructured Materials, Q-MAT,
CESAM, Universit\'e de Li\`ege, B-4000 Sart Tilman, Belgium
}%
\author{W. A. Ortiz}
\author{M. Motta}
\email{m.motta@df.ufscar.br}
\affiliation{ 
Departamento de Física, Universidade Federal de São Carlos, 13565-905 São Carlos, SP, Brazil
}%

\date{\today}

\begin{abstract}
Quantitative magneto-optical imaging of a type-II superconductor thin film cooled under zero, homogeneous, and inhomogeneous applied magnetic fields, indicates that the latter procedure leads to an enhancement of the screening capacity. Such an observation is corroborated by both $B$-independent and $B$-dependent critical state model analyses. Furthermore, repulsive (attractive) vortex-(anti)vortex interactions were found to have a decisive role in the shielding ability, with initial states prepared with vortices resulting in a shorter magnetic flux front penetration depth than those prepared with antivortices. The proposed strategy could be implemented to boost the performance of thin superconducting devices.
\end{abstract}

\maketitle


The ability of type-II superconductors to carry an electric current without dissipation is intrinsically related to how the material can effectively immobilize penetrated quantized flux lines -- the superconducting vortices. In other words, the larger its vortex pinning capacity the higher the critical current density $J_{\text{c}}$.\cite{blatter1994vortices} In the framework of the Bean critical state model,\cite{bean1962magnetization,bean1964magnetization} in which the critical current is independent of the local magnetic flux density $\bm{B}$, the relationship between $J_{\text{c}}$ and the amount of penetrated magnetic flux in a thin film with stripe geometry is:\cite{brandt1993type,zeldov1994magnetization}
\begin{equation} \label{Jc-Bean}
    J_{\text{c}}^{\text{Bean}} = \frac{\pi H}{d \cosh^{-1}\left(\frac{w}{w-p} \right)},
\end{equation}
where $d$ and $w$ are the thickness and the half-width of the film, respectively, $H$ is the intensity of a perpendicularly applied magnetic field and $p$ is the flux front penetration depth measured from the edges.



The enhancement of the pinning capacity is an important quest in developing better superconducting devices for practical applications \cite{foltyn2007materials,gurevich2014challenges,obradors2014coated}. A successful strategy in this regard is to engineer superconducting specimens with artificial pinning center arrays, a series of nanofabricated indentations or inclusions of varied nature spread throughout the material. \cite{civale1991vortex,baert1995composite,martin1997flux,gautam2018probing,zhang2019direct,xu2019ternary,da2020artificial} It has been shown that a graded distribution of holes, or antidots (ADs), emulating the actual vortex distribution in superconducting films can result in $J_{\text{c}}$ values higher than those resulting from a uniform distribution. \cite{misko2012magnetic,motta2013enhanced} In addition, an array of defects arising from the conformal transformation of an annular section of a hexagonal lattice, resembling the Abrikosov vortex lattice,\cite{Abrikosov1957,Kleiner1964} has been predicted to further enhance pinning efficiency.\cite{ray2013strongly} Such conformal crystal structures were achieved experimentally with ADs, confirming an increase in $J_{\text{c}}$. \cite{wang2013enhancing,guenon2013} This kind of defect array preserves features of the local sixfold symmetry of the initial lattice. Moreover, it averts the emergence of flux channeling effects, \cite{menghini2005dendritic,motta2011visualizing} ultimately hindering the occurrence of thermomagnetic flux avalanches which may disrupt superconductivity and be detrimental to the operation of superconducting devices. \cite{vestgaarden2018nucleation,colauto2020controlling} Alternatively, results have shown that Penrose tiling arrays,\cite{misko2005critical,silhanek2006enhanced,kemmler2006commensurability} randomly distributed ADs\cite{wang2016} and disordered hyperuniform AD arrays\cite{lethien2017} may enhance the critical currents over a broad range of applied fields.



Menezes and Souza Silva\cite{menezes2017conformal} showed that a vortex system under a tailored external field spontaneously organizes in a highly inhomogeneous stable array that may be mapped into a hexagonal lattice by a conformal transformation. Later, Menezes \textit{et al.}\cite{menezes2019self} theoretically investigated the vortex landscape of a thin superconducting disk under perpendicular inhomogeneous magnetic fields equivalent to that generated by a concentric current loop. These authors reported that, regardless of the presence of an additional applied homogeneous magnetic field, vortices may self-organize into a variety of defect-free conformal configurations, depending on the thermomagnetic history. Although the behavior of superconducting films under inhomogeneous fields has been studied before, theoretically \cite{milovsevic2002vortex} and experimentally, \cite{govorkov1996,heinrich2004influence} particularly in the context of superconductor/ferromagnetic hybrids,\cite{aladyshkin1999,gillijns2007magnetic,aladyshkin2009nucleation,lyuksyutov2010vortex,BrisboisSciRep2016} an experimental investigation was still lacking, on how field cooling in homogeneous and inhomogeneous
out-of-plane field configurations affects the screening capacity of a macroscopic superconducting film.

In this letter we demonstrate, studying a Nb thin film, that the flux front penetration is affected by different field cooling procedures, being more pronounced when cooling is performed under inhomogeneous fields. Moreover, comparing states prepared either with vortices or antivortices reveals that repulsive vortex-vortex interactions enhance the screening of incoming magnetic flux, while the attraction among vortices and antivortices results in a deeper penetration, indicating a hierarchy on the effective critical current dependent on these interactions and the distribution of previously trapped flux lines.

\begin{figure}
\includegraphics{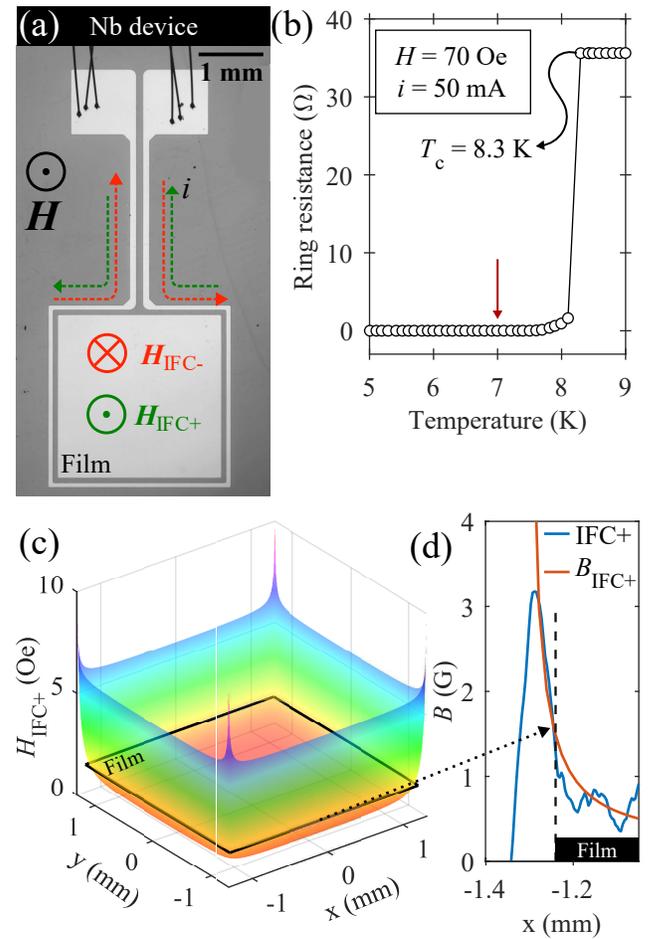}%
\caption{\label{device} (a) Optical image of the Nb device. Current directions and associated magnetic fields are indicated. (b) Temperature-dependent resistance of the ring for $\bm{H}$~=~70 Oe and $i$~=~50~mA. (c) Calculated field distribution $\bm{H_{\text{IFC}+}}(x,y)$ generated by a counter-clockwise 50~mA current applied to the ring. Black square represents the film edges. (d) $\bm{B}(x)$ profile in the film after IFC$+$ procedure obtained from quantitative MOI (blue) plotted alongside calculated $\bm{B_{\text{IFC}+}}$ profile.}%
\end{figure}

To conduct these investigations, we fabricated the Nb superconducting device presented in Fig.~\ref{device}(a). The 200~nm-thick film was grown via dc magnetron sputtering on a Si substrate in a UHV system with base pressure lower than $2\cdot 10^{-8}$ Torr. The device was patterned via optical lithography into a 2.48~mm-wide square film surrounded at a distance of 0.08~mm by a concentric 0.06~mm-wide Nb square ring connected to contact pads allowing for an electric current to pass through. The superconducting critical temperature ($T_c$) at zero dc field and $i$~=~10~mA is 8.5~K. \cite{commentTcMOI}


Measurements of both the temperature in the sample vicinity and the ring resistance showed that a 60~mA current will drive the ring to the normal state and disrupt the thermal equilibrium in the film, prompting the selection of 50~mA as the working current. At such a current and for an applied field of 70~Oe, which leads to a full penetration state in the film, the ring showed a superconducting critical temperature $T_c$~=~8.3~K, as demonstrated in Fig.~\ref{device}(b). Accordingly, it was found that a temperature of $T$~=~7~K was suitable for these measurements to ensure reproducibility.


The magneto-optical imaging (MOI) technique, \cite{vestgaarden2018nucleation} based on the Faraday effect, was used to investigate the flux penetration patterns in the Nb device. The experimental station was equipped with Helmholtz coils to generate a highly uniform magnetic field up to 150~Oe perpendicular to the film. A Bi-substituted yttrium iron garnet film (Bi:YIG) presenting a mostly in-plane spontaneous magnetization was used as Faraday-active indicator.\cite{helseth2002faraday}
Domain walls separating regions presenting different magnetization orientations are known to appear in such magneto-optical materials. These walls are seen as saw-tooth-like lines that are easily displaced, leaving an undesired but unavoidable imprint in the magneto-optical (MO) image. Nonetheless, these domains have negligible influence on the overall flux distribution in the superconducting film.




Moreover, immediate analysis of MO images allows for a qualitative investigation of the flux distribution, as the local brightness is related to the magnitude of the perpendicular flux density. To obtain a quantitative picture, a pixel-by-pixel calibration algorithm \cite{shaw2018quantitative} implemented on MATLAB was used to recover the $\bm{B(x,y)}$ distribution. We also use the plugin StackReg \cite{stackreg} together with ImageJ software \cite{schneider2012nih} to correct for sample drift within a precision of $\pm$2 pixels (or $\pm$8~$\mu$m) in the position of any given image throughout the measurements.




Our main goal was to investigate the effect of different cooling routes on the flux front penetration depth in the Nb film. To achieve that, we prepared the specimen with five different initial states characterized by the cooling: (i) in the absence of a magnetic field, or zero-field cooling (ZFC); (ii) with either a positive or negative uniformly applied magnetic field, or cooling in a homogeneous field (HFC$\pm$); and (iii) using currents flowing through the ring to generate inhomogeneous fields before cooling down the device (IFC$\pm$), as exemplified in Fig.~\ref{device}(a). Then, a positive out-of-plane uniform field $\bm{H}$ was applied to probe the flux penetration in the film. The different field orientations during cooling prepared the film either with a distribution of flux lines in the same direction as the applied magnetic field (vortices, HFC$+$ and IFC$+$) or in the opposite direction (antivortices, HFC$-$ and IFC$-$). This is an important distinction since interactions between vortices are repulsive, \cite{brandt2009vortex} but vortex-antivortex interactions are attractive and may lead to the annihilation of the flux entities. \cite{sardella2009vortex} These interactions have a decisive impact on the penetration dynamics of incoming vortices. \cite{bass1998corrugation,bass1998flux,schwarz2006observation,kramer2011imaging}

Given the different field cooling procedures, one must make sure that images from different measurement runs are comparable. Since flux penetration occurs from the edges to the center of the superconducting films, we compare images in which the effective applied field $\bm{H_{\text{eff}}}$ has the same magnitude at the middle point of the borders. For the ZFC and HFC cases, $\bm{H_{\text{eff}}}=\bm{H}$, however, as current flows through the ring throughout an entire IFC run, in this case $\bm{H_{\text{eff}}}$ is a vectorial sum of the uniform field and the contribution arising from the ring, $\bm{H_{ \text{IFC}\pm}}$. Fig.~\ref{device}(c) presents  $\bm{H_{\text{IFC}+}(x,y)}$ values inside the ring calculated from the Biot-Savart law and reveals that it has a magnitude of 1.5~Oe at the middle of the sample edges. As indicated in Fig.~\ref{device}(d), the flux density in the film after the IFC$+$ procedure matches the behavior of the inhomogeneous field profile (apart from fluctuations due to garnet domains) as one moves away from the borders, confirming our assumption that $\bm{H_{\text{eff}}}=\bm{H}+1.5~\text{Oe}$.

\begin{figure}
\includegraphics{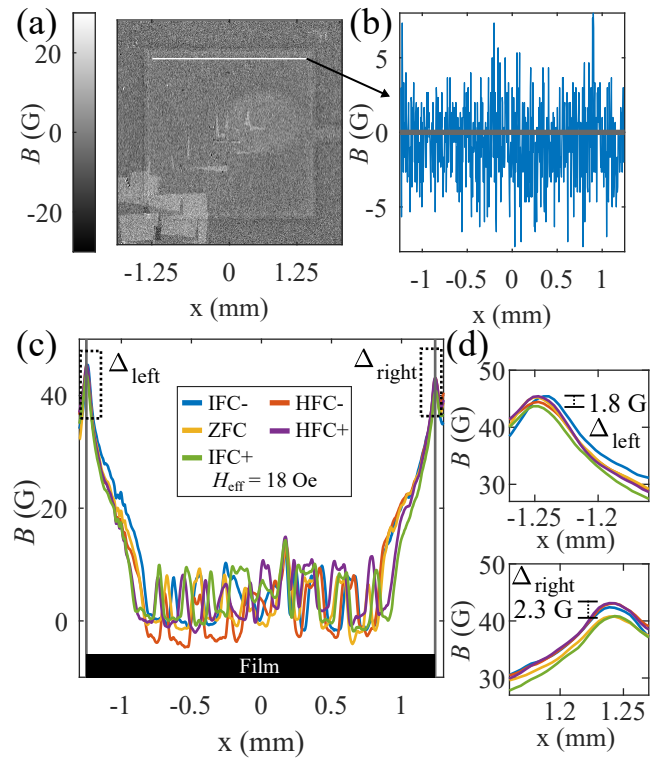}%
\caption{\label{eff-field} (a) Subtraction of HFC$+$ and IFC$+$ field distributions with $\bm{H_{\text{eff}}}$~=~1.5~Oe. (b) Flux distribution along the film edge indicated by white straight line for $\bm{H_{\text{eff}}}$~=~1.5~Oe at 7~K. (c) Flux distribution along the direction indicated by short-dashed line in Fig.~\ref{qualitative} for all cooling procedures with $\bm{H_{\text{eff}}}$~=~18~Oe at 7~K. (d) Details showing $\Delta_{\text{left}}$ and $\Delta_{\text{right}}$ regions highlighted by dashed boxes in panel (c).}%
\end{figure}

\begin{figure*}
\includegraphics{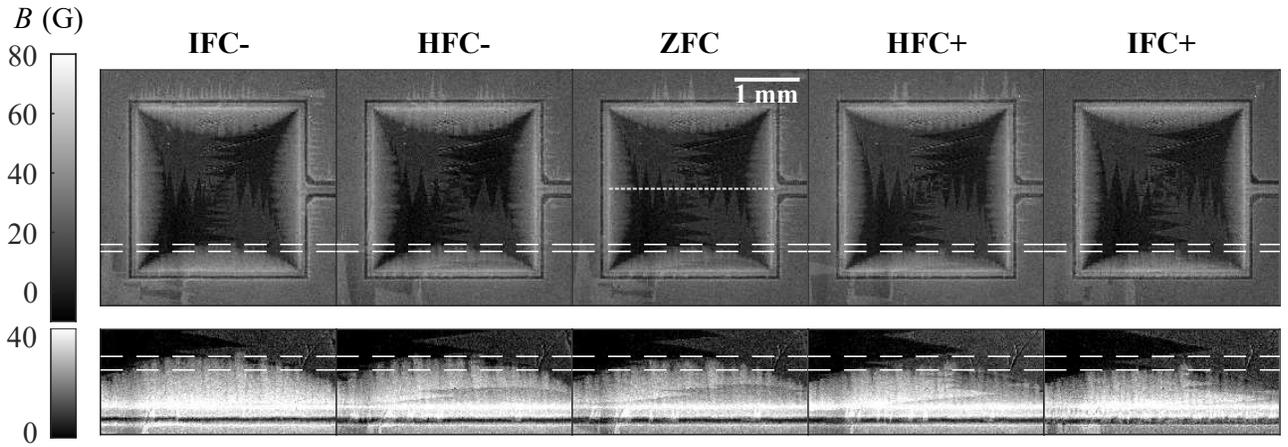}%
\caption{\label{qualitative} Field distribution in the Nb film for $\bm{H_{\text{eff}}}$~=~18~Oe at 7~K and all cooling routes. Bottom panels are zoomed up details of the bottom edges showing the flux front penetration depth for each case. Long-dashed lines are guides to the eye and the short-dashed line in ZFC represents the region of the $B$ profile in Fig.~\ref{eff-field}(c).}
\end{figure*}

One can further attest the validity of this protocol exploring the flux distribution in the film. Fig.~\ref{eff-field}(a) is obtained by performing a pixel-by-pixel subtraction of the local flux density value in the HFC$+$ image from the corresponding point in the IFC$+$ one, both for $\bm{H_{\text{eff}}}$~=~1.5~Oe, i.e., the same field at the edges. When comparing this image to Fig.~\ref{device}(a), one can see that although the current ring is visible, the sample edges are not, indicating that the comparison protocol is valid. Mapping the $\bm{B}$ profile along one of the borders in Fig.~\ref{eff-field}(a) reinforces such fact by showing a $\bm{B}(x)$ distribution around 0~G within the restrictions of our experimental uncertainty that presents a standard deviation of 2.6~G -- see Fig.~\ref{eff-field}(b). Directing our attention to spatial profiles passing through the film center, indicated schematically by the short-dashed line at the ZFC MO image in Fig.~\ref{qualitative}, and investigating all cooling routes with $\bm{H_{\text{eff}}}$~=~18~Oe, one can see in Figs.~\ref{eff-field}(c-d) that a 2D averaging interpolation of the images provides more representative quantitative information, showing that the maximum difference between the fields at the edges for different procedures ($\Delta_{\text{left}}$~=~1.8~G and $\Delta_{\text{right}}$~=~2.3~G, as defined in Fig.~\ref{eff-field}(d)) is again well within our experimental error. In practice, $\Delta$ depends on both $\bm{H_{\text{eff}}}$  and the position chosen to map the field across the sample. Choosing other positions for mapping the field might cause a slight variation on the absolute values of $\bm{B}$ but, regardless where, $\Delta$ is always within the experimental resolution for magnetic fields of our experimental setup. It is important to mention that the intense variations along the field-free central region in the profiles are due to magnetic domains in the indicator, also visible in Fig.~\ref{qualitative}.

If, for example, we take the same $\bm{H_{\text{eff}}}$~=~18~Oe to gauge the penetration patterns after different cooling procedures, the qualitative picture represented in Fig.~\ref{qualitative} emerges. Going from the leftmost (IFC$-$) to the rightmost panel (IFC$+$) a decrease in the flux front penetration depth is observed. This trend is more apparent in the lower row of Fig.~\ref{qualitative} which shows zooms of the bottom edge of the sample. This ordering reveals a hierarchy on the capacity to screen external magnetic fields related to the cooling route, since the left panels represent states prepared with antivortices while the right ones are prepared with vortices. Although our measurements cannot resolve the dynamics of individual vortices, the results suggest that vortex-vortex interactions make it harder for the incoming vortices to penetrate the film as they need to overcome the barrier established by the repulsive potential. Therefore, the positive field cooling procedure may be interpreted as frozen pinned vortices acting as long-range pinning center-like landscapes or magnetic pinning-like distributions. In the HFC case, these frozen vortices are in a uniform distribution, whereas they are in a graded distribution in the IFC case. We could expect that a larger amount of vortices inside the sample would result in a shallower flux front, however, the graded distribution strongly suppresses vortex entry. When, in turn, incoming vortices are faced with antivortices, the attractive potential facilitates vortex penetration since annihilation processes may be allowing them to further penetrate the film, resulting in the overall effect observed in Fig.~\ref{qualitative}. 


Moreover, one notices that the extreme cases in Fig.~\ref{qualitative} are IFC (``$-$" on the left and ``$+$" on the right extremity) indicating that the initial flux distribution arising from cooling under inhomogeneous fields has a more pronounced effect on the flux front penetration depth than the homogeneous counterparts. These observations are in line with previous results showing that non-uniform pinning center distributions enhance the screening capacity of superconductors \cite{motta2013enhanced,wang2013enhancing} and that inhomogeneous magnetic fields can be used to create optimal vortex arrangements to improve pinning. \cite{menezes2019self} Such analysis is independent of any numerical data treatment and can be made directly from the raw intensity distribution obtained from MOI, dramatically diminishing the importance of the visible garnet domains in the results. 

Turning to the Bean model, the shorter flux penetration depth observed as one moves towards the IFC$+$ case implies a higher effective current density, i.e., a higher screening current flowing through the film at the same effective field. This is shown in Table \ref{table} where the values of $p$ were measured directly from Fig.~\ref{qualitative} within a 6-pixel uncertainty. Besides the absolute values estimated for $J_{\text{eff}}^{\text{Bean}}$ from Eq.~(\ref{Jc-Bean}), the result is also presented in numbers relative to the ZFC case, indicating the percentage variation observed for each procedure. 

\begin{table}
\caption{\label{table} Comparison between current density variation obtained from the Bean and Kim models.}
\begin{ruledtabular}
\begin{tabular}{c|ccc|c}
              & \multicolumn{3}{c|}{Bean model} & \multicolumn{1}{c}{Kim model} \\
              Procedure & \begin{tabular}[c]{@{}c@{}}$p \pm 0.024$\\ (mm)\end{tabular} & \begin{tabular}[c]{@{}c@{}}$J_{\text{eff}}^{\text{Bean}}$\\ (10$^5$ A/cm$^2$)\end{tabular} & \begin{tabular}[c]{@{}c@{}}$J_{\text{eff}}^{\text{Bean}}$\\ (\%)\end{tabular} &  \begin{tabular}[c]{@{}c@{}}$J_{\text{eff}}^{\text{Kim}}$\\ (\%)\end{tabular} \\ \hline
\textbf{IFC-} & 0.383  & 24.6 $\pm$ 1.0 & 94.2   & 95.6  \\
\textbf{HFC-} & 0.375  & 25.0 $\pm$ 1.1  & 95.6   & 97.5  \\
\textbf{ZFC}  & 0.350  & 26.2 $\pm$ 1.2 & 100    & 100    \\
\textbf{HFC+} & 0.300  & 28.9 $\pm$ 1.4 & 110.3  & 103.4 \\
\textbf{IFC+} & 0.288  & 29.6 $\pm$ 1.5 & 113.2  &  107.3
\end{tabular}
\end{ruledtabular}
\end{table}

\begin{figure}
\includegraphics{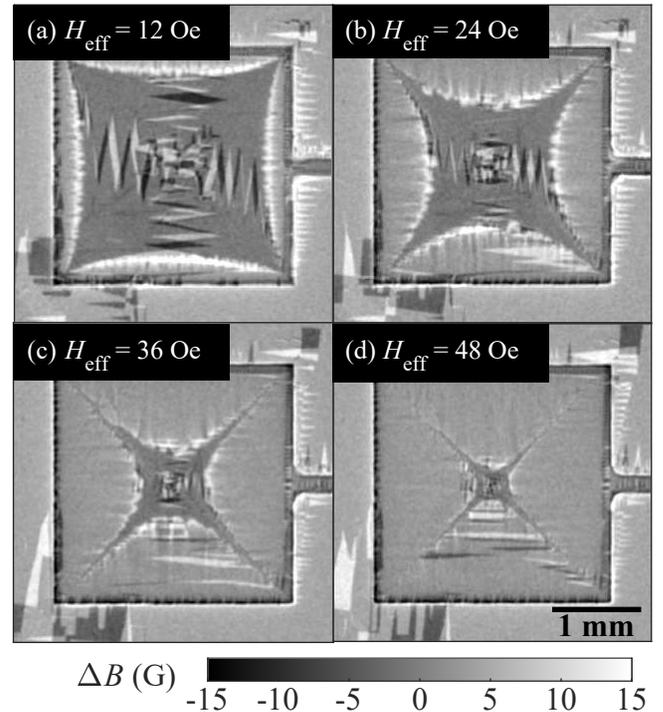}%
\caption{\label{fluxVarHeff} Subtraction of the IFC$+$ image from the corresponding IFC$-$ for different $\bm{H_{\text{eff}}}$: (a) 12~Oe, (b) 24~Oe, (c) 36~Oe, and (d) 48~Oe. Brighter pixels indicate a positive contrast.}
\end{figure}

The same behavior can also be observed for different effective applied fields. Subtracting IFC$+$ images from IFC$-$ ones, in a procedure similar to that presented in Figure~\ref{eff-field}(a), the resulting contrast highlights the difference in the flux penetration. For instance, if the flux penetration of image ``A" is deeper than that of image ``B", subtracting B from A would result in a positive (bright) contrast at the flux front. This is precisely what is observed in Fig.~\ref{fluxVarHeff} for $\bm{H_{\text{eff}}}$~=~12~Oe, 24~Oe, 36~Oe, and 48 Oe, fading out for higher effective fields. Therefore, the IFC$+$ case always presents the shortest flux front penetration depth and, consequently, the higher screening capacity.


A different analysis based on the Kim model~\cite{kim1962critical,chen1989kim} was done to account for a $J_{\text{c}}(\bm{B})$ dependency that would lead to differences in both flux penetration and current distribution patterns in superconducting films~\cite{mcdonald1996theory}. This dependency was recently shown necessary to explain experimental observations on the triggering of flux avalanches in Nb films~\cite{jiang2020selective}. Therefore, an analysis beyond the Bean model is desirable for this data.



From average $B(y)$ profiles considering 5 rows along the center of the film, current density distributions were obtained by means of numerical calculations described in Ref.~\citenum{mcdonald1996theory}. The results are normalized by the unknown critical current density at zero applied field $J_{\text{c0}}$. Then, despite the influence of garnet domains, defining the effective current density $J_{\text{eff}}^{\text{Kim}}$ as the absolute current value in the maximum flux front penetration depth of the IFC$+$ case allows one to firmly state that, where there is flux penetrated, $J_{\text{eff}}^{\text{Kim}}$ is higher as you move from left to right in Fig.~\ref{qualitative}, as represented in Table~\ref{table} for $\bm{H_{\text{eff}}}$~=~18~Oe.

\begin{figure}
\includegraphics{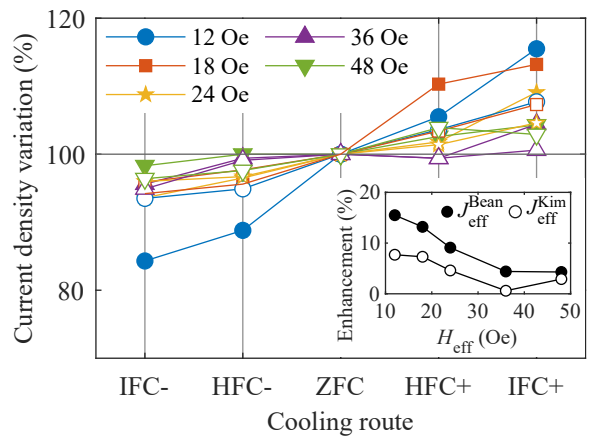}%
\caption{\label{varHeff} Schematic representation of the current density variation for different field cooling procedures and $\bm{H_{\text{eff}}}$. Inset shows the evolution of current density enhancement with $\bm{H_{\text{eff}}}$. In both panels $J_{\text{eff}}^{\text{Bean}}$ (closed symbols) and $J_{\text{eff}}^{\text{Kim}}$ (open symbols) analysis are represented.}
\end{figure}

Evaluating $J_{\text{eff}}^{\text{Kim}}$ for different $\bm{H_{\text{eff}}}$ allows us to reach the same conclusion, i.e there is a consistent enhancement of the screening capacity in states prepared with vortices, that is maximized for a field cooling performed with inhomogeneous fields. Fig.~\ref{varHeff} summarizes this observation highlighting an upward trend in current variation from the left to the right side in both analysis. The inset shows the maximum current enhancement for different $\bm{H_{\text{eff}}}$; the decrease observed in $J_{\text{eff}}^{\text{Bean}}$ matches the tendency in Fig.~\ref{fluxVarHeff}. Also, the apparent smaller enhancement for $J_{\text{eff}}^{\text{Kim}}$ might result from the $J_{\text{c0}}$ normalization, which must probably has different values for the different cooling routes.

In conclusion, we have fabricated a superconducting device that provides controllable applications of inhomogeneous magnetic fields in a Nb film. Quantitative MOI revealed the influence of different cooling routes on the flux front penetration depth. Both Bean and Kim models indicate that cooling procedures under inhomogeneous magnetic fields have the strongest impact on the effective shielding current flowing throughout the superconductor.
Even though the spatial resolution of our MOI station does not allow for a
statement to be made on the vortex arrangements in comparison to those predicted by Menezes \textit{et al.}\cite{menezes2019self}, our findings indicate that cooling in an inhomogeneous field is a viable route to enhance its effective critical current. Moreover, the MO images show that when the film was initially prepared with states permeated by antivortices, the flux penetration was deeper. This fact hints at a hierarchy on the screening capacity dependent on the nature of the interactions of the incoming vortices with previously pinned flux lines, i.e., repulsive vortex-vortex interactions hamper flux penetration, which is translated into a higher effective screening current. Therefore, not only the interaction mechanism is recognized as an important ingredient to comprehend these results, but also the frozen vortex distribution throughout the material. Our findings may also be extended to bulky materials, as hinted by results described by Morita \textit{et al.}~\cite{morita1992magnets} for a high-$T_c$ specimen cooled in a homogenous field. Further examination using a different experimental approach which reaches individual vortex resolution is necessary to prove the conformal distribution of vortices. Additionally, we foresee that the initial state may also influence the threshold field to trigger flux avalanches.


\begin{acknowledgments}
The authors thank C. C. de Souza Silva and L. R. E. Cabral for the enlightening discussions, Laboratório de Conformação Nanométrica (LCN/IF/UFRGS) for the Nb film growth, and Laboratório de Microfabricação (LMF/LNNano/CNPEM) for the litography. The authors also acknowledge financial support from Brazilian agencies: Coordenação de Aperfeiçoamento de Pessoal de Nível Superior -- Brasil (CAPES) -- Finance Code 001; National Council for Scientific and Technological Development (CNPq); and S\~ao Paulo Research Foundation (FAPESP).
\end{acknowledgments}

\begin{section}*{Data Availability}
The data that support the findings of this study are available from the corresponding author upon reasonable request.
\end{section}

\bibliography{refs}

\begin{thebibliography}{57}
\bibitem [{com()}]{commentTcMOI}%
  \BibitemOpen
  \href@noop {} {\enquote {\bibinfo {title} {{T}here is a difference between
  {$T_c$} obtained through measurements in a {Q}uantum {D}esign {SQUID}
  magnetometer ($\sim$9~{K}) and our {MOI} setup (8.5~{K}). {I}t appears mainly
  as a consequence of a non-optimum thermal contact between the sample and the
  cold-finger type cryostat. {D}espite that, this difference does not affect
  the results shown here.}}\ }\BibitemShut {NoStop}%
\end{thebibliography}

\end{document}